\documentclass[]{spie}
\usepackage[]{graphicx}
\usepackage{amssymb}

\pdfoutput=1

\title{ACTPol: A polarization-sensitive receiver \\
for the Atacama Cosmology Telescope}

\author{M. D. Niemack$^1$, P. A. R. Ade$^2$, J. Aguirre$^3$, F. Barrientos$^4$, J. A. Beall$^1$, J. R. Bond$^5$, \\
J. Britton$^1$, H. M. Cho$^1$, S. Das$^6$, M. J. Devlin$^3$, S. Dicker$^3$, J. Dunkley$^7$, R. D\"{u}nner$^4$, \\
J. W. Fowler$^8$, A. Hajian$^5$, M. Halpern$^{9}$, M. Hasselfield$^{9}$, G. C. Hilton$^1$, M. Hilton$^{10}$, \\
J. Hubmayr$^1$, J. P. Hughes$^{11}$, L. Infante$^{4}$, K. D. Irwin$^1$, N. Jarosik$^8$, J. Klein$^3$, A. Kosowsky$^{12}$,\\
T. A. Marriage$^{13}$, J. McMahon$^{14}$, F. Menanteau$^{11}$, K. Moodley$^{10}$, J. P. Nibarger$^1$, M. R. Nolta$^{5}$, \\
L. A. Page$^8$, B. Partridge$^{15}$, E. D. Reese$^3$, J. Sievers$^5$, D. N. Spergel$^{13}$, S. T. Staggs$^8$, \\
R. Thornton$^{16}$, C. Tucker$^2$, E. Wollack$^{17}$, K. W. Yoon$^1$
\skiplinehalf
\small{
$^1$ National Institute of Standards and Technology, 325 Broadway MC 817.03, Boulder, CO 80305, USA \\
$^2$ School of Physics and Astronomy, Cardiff Univ., The Parade, Cardiff, Wales, UK CF24 3AA \\
$^3$ Department of Physics and Astronomy, Univ. of Pennsylvania, 209 South 33rd Street, Philadelphia, PA, USA 19104 \\
$^4$ Departamento de Astronom\'{i}a y Astrof\'{i}sica, Pontific\'{i}a Univ. Cat\'{o}lica, Casilla 306, Santiago 22, Chile \\
$^5$ Canadian Institute for Theoretical Astrophysics, Univ. of Toronto, Toronto, ON, Canada M5S 3H8 \\
$^6$ Berkeley Center for Cosmological Physics, LBL and Dept. of Physics, Univ. of California, Berkeley, CA, USA 94720 \\
$^7$ Department of Astrophysics, Oxford University, Oxford, UK OX1 3RH \\
$^8$ Joseph Henry Laboratories of Physics, Jadwin Hall, Princeton University, Princeton, NJ, USA 08544 \\
$^{9}$ Department of Physics and Astronomy, University of British Columbia, Vancouver, BC, Canada V6T 1Z4 \\
$^{10}$ School of Mathematical Sciences, University of KwaZulu-Natal, Durban, 4041, South Africa \\
$^{11}$ Department of Physics and Astronomy, Rutgers, The State University of New Jersey, Piscataway, NJ USA 08854 \\
$^{12}$ Department of Physics and Astronomy, University of Pittsburgh, Pittsburgh, PA, USA 15260 \\
$^{13}$ Department of Astrophysical Sciences, Peyton Hall, Princeton University, Princeton, NJ USA 08544 \\
$^{14}$ Physics Department, University of Michigan, 450 Church Street, Ann Arbor, Michigan 48109 \\
$^{15}$ Department of Physics and Astronomy, Haverford College, Haverford, PA, USA 19041 \\
$^{16}$ Department of Physics , West Chester University of Pennsylvania, West Chester, PA, USA 19383 \\
$^{17}$ NASA/Goddard Space Flight Center, Code 553/665, Greenbelt, MD, USA 20771 \\
}}

\authorinfo{Send correspondence to niemack@nist.gov.}

\begin{document}
  \maketitle

\begin{abstract}
The six-meter Atacama Cosmology Telescope (ACT) in Chile was built to measure the cosmic microwave background (CMB) at arcminute angular scales. We are building a new polarization sensitive receiver for ACT (ACTPol). ACTPol will characterize the gravitational lensing of the CMB and aims to constrain the sum of the neutrino masses with $\sim0.05$ eV precision, the running of the spectral index of inflation-induced fluctuations, and the primordial helium
abundance to better than 1~\%. Our observing fields will overlap with the SDSS BOSS survey at optical wavelengths, enabling a variety of cross-correlation science, including studies of the growth of cosmic structure from Sunyaev-Zel'dovich observations of clusters of galaxies as well as independent constraints on the sum of the neutrino masses. We describe the science objectives and the initial receiver design.
\end{abstract}


%%%%%%%%%%%%%%%%%%%%%%%%%%%%%%%%%%%%%%%%%%%%%%%%%%%%%%%%%%%%%
\section{INTRODUCTION}
\label{sec:intro}

The Atacama Cosmology Telescope (ACT) has been used since 2008 to observe temperature anisotropies in the Cosmic Microwave Background (CMB) between wavelengths of one to two millimeters \cite{fowler/etal:2007}. These observations led to improved measurements of the CMB temperature anisotropy power spectrum \cite{fowler/etal:2010} on arcminute angular scales and detections of galaxy clusters via the Sunyaev-Zel'dovich (SZ) effect \cite{hincks/etal:2009}. Multi-frequency data reduction and analysis of point sources, blind SZ detections, and cosmological constraints are underway, while observations with ACT continue.

We are developing a new dual-frequency (150 GHz and 220 GHz) polarization sensitive receiver (ACTPol) to be deployed on ACT in 2013. In addition to being polarization sensitive, it is projected to substantially improve upon the temperature sensitivity of the current receiver.  These improvements will enable an expansion of the current ACT science goals, which are based on a range of observations that go well beyond the CMB power spectra, and combine to provide a built-in set of cross-checks.  ACTPol will focus on a four-pronged science program:
\begin{enumerate}
\item { Measure the intrinsic temperature and polarization anisotropy at high-multipoles ($\ell$) to probe the spectral index of inflation, the primordial helium abundance, and neutrino properties.}
\item { Measure the gravitational lensing of the CMB in temperature and polarization to constrain early dark energy and the sum of the neutrino masses.}
\item  { Correlate and compare the CMB with lower redshift optical, infrared, and X-ray surveys to achieve a variety of science goals, including independent constraints on the sum of the neutrino masses.}
\item { Find clusters of galaxies through their Sunyaev-Zel'dovich effect and combine them with other surveys to study the growth of structure.}
\end{enumerate}

These four measurement objectives and science goals are described in more detail in $\S$\ref{sec:science} where we also present predicted results based on the target ACTPol performance. The ACTPol observing strategy, foreground emission, and calibration are discussed in $\S$\ref{sec:obs}, the ACT design is reviewed in $\S$\ref{sec:telescope}, and an initial ACTPol receiver design and sensitivity forecast are introduced in $\S$\ref{sec:receiver}.

\section{SCIENCE GOALS}
\label{sec:science}

\begin{figure}
   \begin{center}
   \includegraphics[width=3.3in]{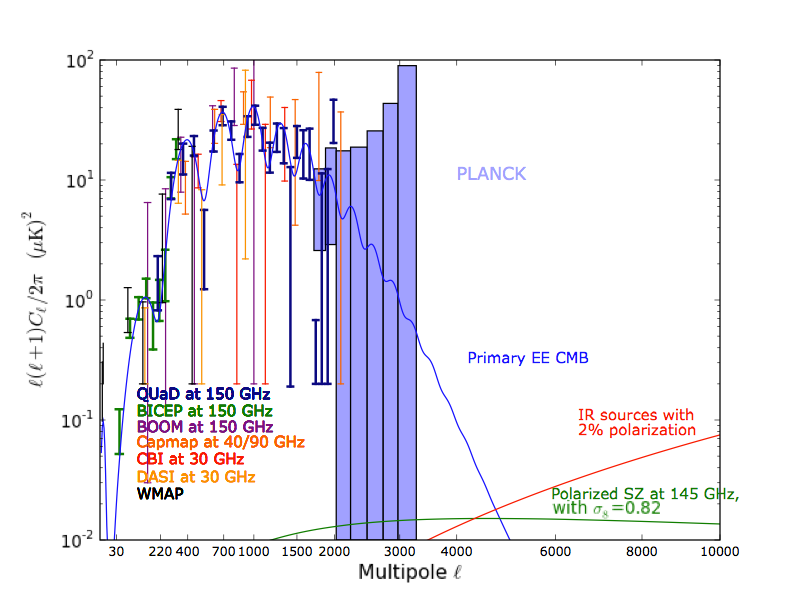}
   \includegraphics[width=3.3in]{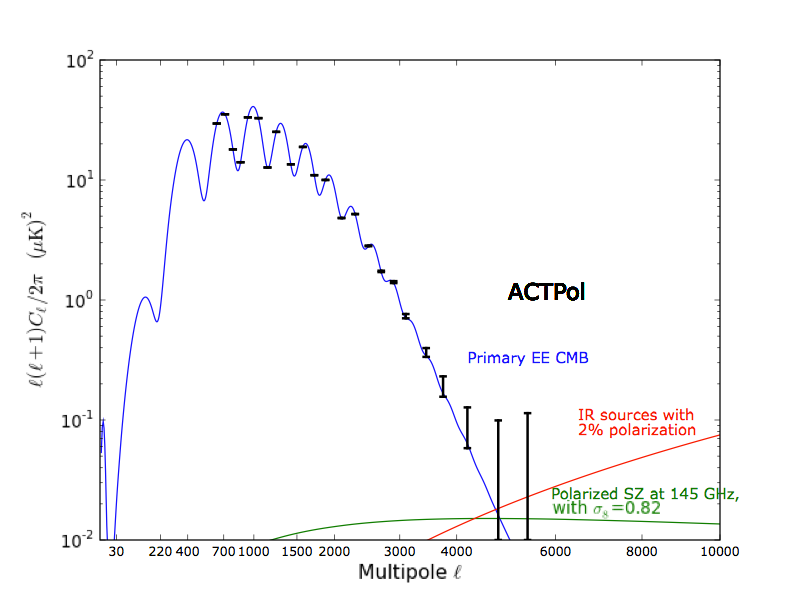}
   \end{center}
\caption{\small {\it Left:} The current state of measurements of the E-mode polarization (EE) power spectrum
plotted over the best fit cosmological model. Data are from QUaD \cite{brown/etal:2009}, BICEP
\cite{chiang/etal:2009}, Boomerang \cite{montroy06}, CAPMAP \cite{bischoff08}, CBI \cite{sievers07}, DASI \cite{kovac/etal:2002}, and WMAP\cite{nolta/etal:2009}.  The projected Planck error bars (from the ``Planck Bluebook") are shown as blue boxes. The foregrounds
are conservative estimates for infrared point sources and a 1~\% net polarization
of the SZ effect. {\it Right:} Projection of the ACTPol errors for the deep fields (150 deg$^2$) based on the target instrument sensitivity described in $\S$\ref{sec:receiver} and 1300 hours of observations by use of the Knox formula \cite{knox95}, which ignores foregrounds, systematic errors, and extra noise from the atmosphere. We do not project errors at $\ell < 500$ because the Knox formula is inaccurate at large scales in the presence of atmospheric fluctuations.}
\label{fig:pol}
\end{figure}

\subsection{Measure the Intrinsic Temperature and Polarization Anisotropies at $\ell \gtrsim 2000$}

The temperature anisotropy may be divided roughly  into large ($\ell \lesssim 2000$) and small ($\ell \gtrsim 2000$) angular scales.
At large angular scales accessible to the WMAP and Planck\footnote{Planck uncertainties from: http://www.rssd.esa.int/Planck} satellite telescopes,
the anisotropy is a direct probe of the response of the CMB to perturbations laid down in the early Universe as seen at a redshift of $z \sim 1000$.  At smaller angular scales nonlinear physics (structure formation) dominates, and the foreground contamination by point sources such as radio and starforming galaxies becomes significant. Measuring the transition between these two regimes in both temperature and polarization will improve constraints on the standard cosmological model.

\begin{figure}
\begin{center}
 \includegraphics[width=5in]{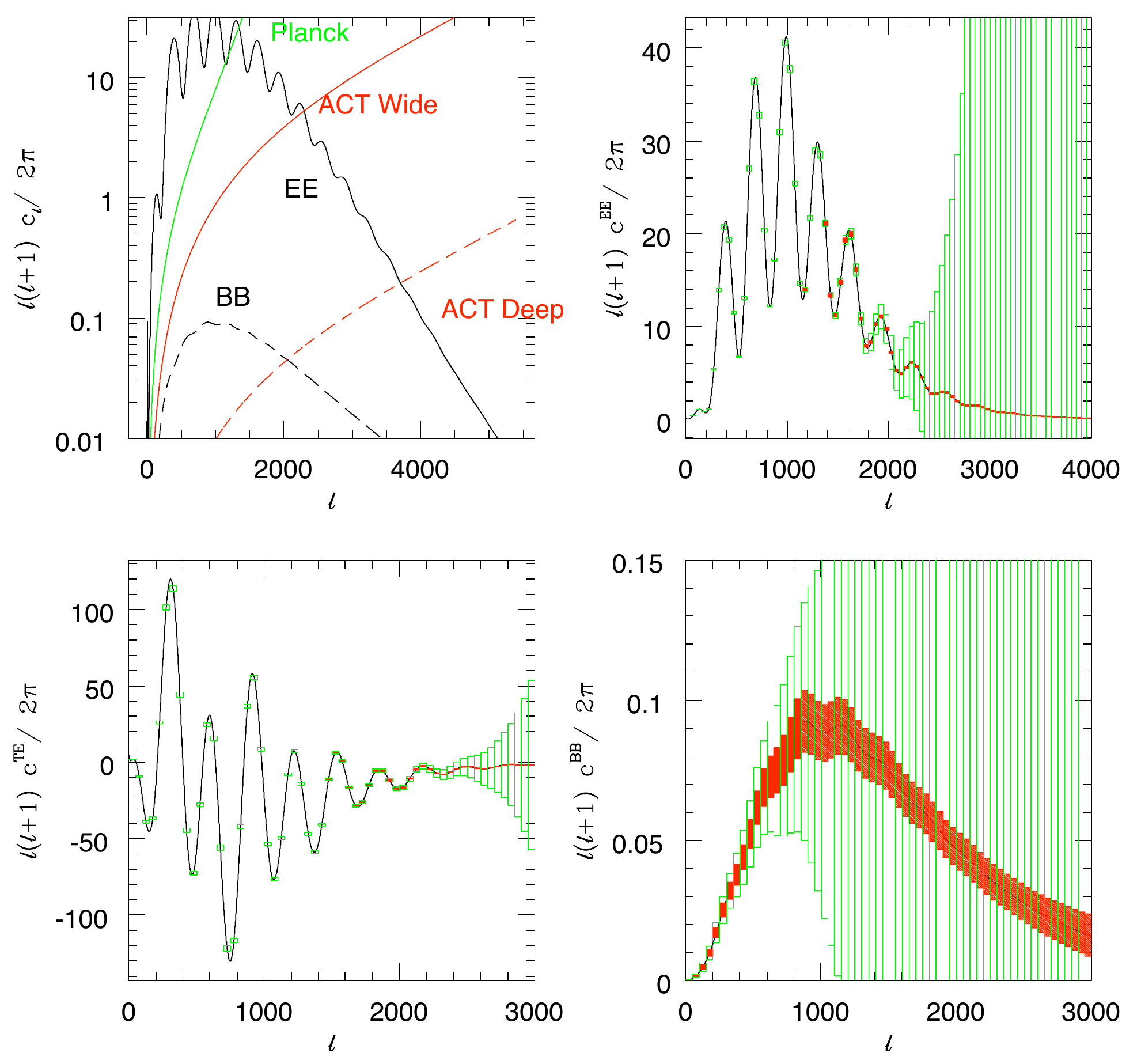}
\caption{The upper left panel shows the {\it statistical} noise per multipole for Planck, ACTPol Wide, and ACTPol Deep ($\S$\ref{sec:obs}) and the predicted EE and BB power spectra. The BB spectrum is due to lensing. The other three panels show the EE, BB, and TE spectra with target ACTPol errors (filled red boxes, which are small on a linear scale except in the BB spectrum; Figure~\ref{fig:pol} shows the EE spectrum on a logarithmic scale) and Planck errors (open green boxes).  At low $\ell$, when the Planck errors are smaller, only they (open green boxes) are shown.  In the BB spectrum the red boxes at $\ell < 850$ are for the ACTPol wide survey ($\S$\ref{sec:obs}), while the red boxes at $\ell > 850$ are for the ACTPol deep survey. The latter three panels are binned with $\Delta \ell = 50$. Atmospheric contamination may preclude ACTPol measurements below $\ell\sim500$; however, the atmosphere is not polarized at these frequencies, which could enable polarization measurements at lower $\ell$ than temperature measurements. The y-axes are in $\mu$K$^2$.}
\label{fig:sisnoise}
\end{center}
\end{figure}

Our planned ACTPol surveys (\S\ref{sec:obs}) are designed to continue characterizing the temperature (TT) foregrounds out
to $\ell \sim 8,000$ and to measure the E-mode polarization (EE), temperature to E-mode (TE), and B-mode polarization (BB) power spectra with high signal-to-noise ratio (see Figures \ref{fig:pol} and \ref{fig:sisnoise}) out to $\ell \gtrsim 3000$. These data will allow us to probe early-Universe physics, reduce errors on cosmological parameters, measure
the number of neutrino species, and determine the helium abundance (Table~1). We address the science related to the various power spectra below.   Although the results are derived from the combined analysis of all spectra, we concentrate on TT and EE in this section, and focus on BB when we discuss lensing (\S\ref{sec:lensing}).

\paragraph{Inflation and early Universe physics.}
The small-scale temperature anisotropy is important for testing models of the early Universe.
The most favored model is inflation \cite{guth:1981,linde:1982, albrecht/steinhardt:1982,
sato:1981}, though there are variants and alternatives.  Assuming adiabatic and Gaussian fluctuations, there are three early-Universe  parameters one can get directly from the CMB:
the scalar spectral index, $n_s$, or overall slope of the angular power
spectrum after removing the acoustic oscillations \cite{mukhanov/chibisov:1981,hawking:1982,
guth/pi:1982, starobinsky:1982, bardeen/steinhardt/turner:1983};  the running
of the spectral index with wavenumber $k$ or $\alpha\equiv d\,n_s/d{\rm ln}\,k$
\cite{kosowsky/turner:1995}; and the tensor to scalar ratio, $r$, which describes
the degree to which the Universe is suffused with a cosmic background
of gravitational waves.  Of these, ACTPol will focus on constraining $n_s$ and $\alpha$ by measuring the $\ell>500$ E-mode polarization. In principle ACTPol can measure the low-$\ell$ gravitational wave B-mode spectum, but we believe the low-$\ell$ polarization measurements are best done with optimized dedicated instruments.  Although inflation-related tensor modes are directly manifest at low $\ell$, high-$\ell$ measurements can also constrain them through their covariance with other parameters, most notably $n_s$.

All indications are that $n_s$ is slightly less than unity, which is in line with the
predictions of simple models of inflation, $n_s\approx0.95$. Recent work by the QUaD collaboration \cite{brown/etal:2009} gives an example of what can be done by adding just $\ell<2000$ CMB data (ACBAR+QUaD) to the WMAP results. They find $n_s=0.96\pm0.012$ and $\alpha=-0.028\pm0.018$.

High-resolution and high-sensitivity E-mode measurements offer a new way to measure $n_s$ and
$\alpha$.  Even though the EE spectrum at $\ell\sim3000$ is one-sixth as large in temperature as the TT spectrum, the polarized point source contamination is far less (see $\S$\ref{sec:obs}).  Thus, EE measurements offer a relatively point-source-contamination-free way to measure the key inflation parameters, $n_s$ and $\alpha$. Through the EE spectrum, ACTPol should be able to directly probe the primordial power spectrum out to $\ell\sim3500$.

\begin{table}
%[htdp]
\begin{center}
\begin{tabular}{|c|c|c|c|c|}
\hline
\hline
&WMAP5 &WMAP9 + ACTPol & Planck & Planck + ACTPol \\
\hline
$\Omega_b h^2$ &$6\times10^{-4}$& $1.9\times  10^{-4}$& $2 \times 10^{-4}$&$ 1.3\times 10^{-4}$\\
$\Omega_m h^2$ &$6\times10^{-3}$ &$3.7 \times 10^{-3}$ &$2.2 \times 10^{-3} $&$1.7 \times 10^{-3}$\\
$n_s$ &0.014&0.008 & 0.007&0.006 \\
%$\ln \theta_A$&$ 3\times 10^{-3}$ &$1.2\times 10^{-3}$&$4 \times 10^{-4}$ &$5 \times 10^{-4}$&$3
%\times 10^{-4} $\\
$m_\nu$ & - & 0.15& 0.1&0.06 \\
$\tau$&0.017 &0.010 & 0.004&0.004 \\
$Y_{He}$&- &0.007 & 0.01 &0.005 \\
\hline
\hline
\end{tabular}
\caption{This table shows projected 68~\% confidence interval constraints in an 8-parameter model (baryon density, matter density, $n_s$, neutrino mass (in eV), optical depth, helium abundance, Hubble expansion factor, amplitude) from the ACTPol data with WMAP and Planck based on a Fisher matrix analysis without systematic errors.  For a 7-parameter model
(baryon density, matter density, $n_s$, optical depth, Hubble expansion factor, amplitude, $\alpha$) with Planck+ACTPol
the constraint on $\alpha$ is 0.004.
For comparison, current constraints from WMAP5 are shown in which only 6 parameters are fit. The WMAP data set after 9 years is named WMAP9. }
\end{center}
\end{table}

\paragraph{Non-Gaussianity.} In the standard model of cosmology, the fluctuations are Gaussian. It is widely believed that if the model is incomplete or incorrect the first hints
will come through the detection of non-Gaussianity \cite{komatsu:2010}. To date there are only upper limits on non-Gaussianity. As one example of physics beyond the standard model, we studied the detectability of cosmic strings through their imprint on the CMB \cite{fraisse/etal:2008} in the presence of secondary anisotropies and found that ACTPol could place improved limits of G$\mu<10^{-7}$ (in Planck units) on the string tension.

\paragraph{Helium abundance, neutrinos, relativistic species, and the standard model.}
The high-$\ell$ EE spectrum is sensitive to the primordial helium abundance because
the associated high-$\ell$ modes enter the horizon before the helium fully recombines
\cite{ichikawa08,iocco09}.
We aim to measure the primordial helium abundance, $Y_{He}$, to better than 1~\%,
independent of systematic effects in the astronomical determination of the abundance,
 which will allow precise measurements of the baryon density and thus the spectral index\footnote{Currently, the parameter with the greatest degeneracy with $n_s$ is $\Omega_bh^2$. The optical depth, $\tau$, is known well enough that the $n_s-\tau$ degeneracy is broken.}.
Our measurement of helium abundance, shown in Table~1, should have the same errors as those quoted for measurements of primordial helium abundance from white dwarfs \cite{fukugita/kawasaki:2006}.  The combination of precise measurements
of helium abundances and $\Omega_b h^2$ will enable powerful constraints on neutrino properties and early-Universe physics.  ACTPol aims to constrain the sum of the neutrino masses
in multiple ways, with projected accuracies between 0.05 and 0.07 eV ($\S$\ref{sec:lensing}), a substantial improvement on the recent WMAP constraints of $<1.3$ eV \cite{dunkley/etal:2009}.
Measurements of atmospheric neutrino oscillations imply $\Delta m_\nu \simeq 0.05$~eV \cite{adamson/etal:2008}. Thus, ACTPol should measure the sum of the neutrino masses, and if the sum is found to be  $\lesssim 0.1$~eV, this measurement would rule out the possibility of an inverted neutrino mass hierarchy \cite{jimenez/etal:2010}.

A high-$\ell$ measurement of the EE spectrum will allow significantly improved constraints on the effective number of neutrino species, which in the standard model is $N_{\rm{eff}} = 3.04$. Additional relativistic species (e.g., gravitons, Goldstone bosons)
could contribute to the radiation density or non-standard neutrino interactions could exist that would alter $N_{\rm{eff}}$, either of which would induce a distinct phase shift in the positions of the acoustic peaks  \cite{bashinsky/seljak:2004}. Current constraints from WMAP and other cosmological data give $N_{\rm{eff}} = 4.3\pm0.9$ \cite{Komatsu/etal:2010}. ACTPol will probe physics beyond the standard model by improving this constraint by more than a factor of two.

\subsection {CMB Lensing and B-mode Polarization\label{sec:lensing}}

At high $\ell$, the B-mode polarization is produced by gravitational lensing of the E-modes
\cite{zaldarriaga/seljak:1998}.  The lensing power spectrum can give an excellent measure of the neutrino mass sum, spatial curvature, and ``early dark energy" \cite{deputter/zahn/linder:2009}.  CMB lensing provides the best way to study the nature of dark energy at early times, because the lensing kernel probes a wide range of redshifts that peaks at $z \sim 2$-4, while low-redshift cosmological probes, including optical galaxy lensing measurements, are sensitive to cosmology at $z \lesssim 1$.  ACTPol is designed to measure the lensing in both temperature and polarization.

ACTPol's measurements of the CMB lensing power spectrum can provide an accurate measurement of the amplitude of matter fluctuations at $z \sim 2$-4.  This measurement, in turn, can provide a sensitive determination of neutrino mass of $\sim0.07$~eV, close to the observed mass splitting (see Table~1); and complement weak lensing surveys that trace the mass fluctuations in mass sheets at $z \lesssim 1$.  By providing a high-$z$ anchor for planned dark energy surveys, ACTPol will improve measurements of dark-energy effects on structure growth.

ACTPol should be able to measure lensing to a high enough precision to ``delens", or reconstruct the true CMB fluctuation power spectrum.  It could thus improve the S/N on
large-angular-scale, primordial B-mode measurements by a factor of $\sim 1.8$ \cite{smith2008}. By combining CMB lensing data with measurements from the surface of last scattering to break the geometric degeneracy \cite{stompor99}, WMAP9 + ACTPol should be able to measure the curvature of the universe to better than 0.5~\% by use of CMB data alone.

By cross-correlating the lensing of the CMB with the variance in the fluctuations in the
Ly$\alpha$ forest, we will directly probe the relationship between matter and gas in the Ly$\alpha$ forest.  Vallinotto {\it et al.} \cite{vallinotto/etal:2009} estimate a
S/N = 20 cross-correlation signal for ACTPol with BOSS.  This  measurement leads to a 5~\% determination of the amplitude of the Ly$\alpha$ power spectrum, which leads to another measurement of neutrino mass with $\sigma_\nu \sim 0.05$~eV.	

The combination of WMAP and the SDSS\footnote{SDSS:  The Sloan Digital Sky Survey.  http://www.sdss.org/} Luminous Red Galaxy (LRG) survey ($z \sim 0.3$) has already placed powerful limits on cosmological parameters \cite{percival/etal:2009,reid/etal:2009}.  The BOSS\footnote{BOSS:  The Baryonic Oscillation Spectroscopic Survey.  http://cosmology.lbl.gov/BOSS/} LRG sample will extend the range, providing an accurate determination of the galaxy power spectrum for $z = 0.3$-0.7.
Acquaviva {\it et al.} \cite{acquaviva/etal:2009} have estimated a S/N of 11, 25 and 40 for cross correlations of ACTPol with SDSS DR7 LRG, BOSS low-redshift LRG, and BOSS LRG samples.  This translates to another independent
measurement of neutrino mass with $\sigma_\nu\sim0.05$~eV, an improvement over the ACTPol-only constraint,  and provides yet another method of assessing the effects of dark energy on
structure growth \cite{acquaviva/etal:2009}.

\subsection {SZ Clusters, an Unbiased Search}
One well studied perturbation of the CMB at small angular scales is the Sunyaev-Zel'dovich (SZ) effect, in which the hot electrons ($10^7$~K-$10^8$~K, $\sim5$\,keV) in galaxy clusters reveal their presence by scattering the CMB with a characteristic frequency signature \cite{Sunyaev_Zeldovich:1980}. The process of finding hundreds to thousands of clusters in large blind and unbiased surveys has already been started by the SPT \cite{staniszewski/etal:2009} and ACT \cite{hincks/etal:2009} groups. These are the two best telescopes in the world for these studies. ACT enjoys the ability to overlap with many optical and radio observations of clusters made by telescopes in both the northern and southern hemispheres. We aim to find $\sim1000$ clusters in the ACTPol surveys \cite{sehgal/etal:2010,Tinker/etal:2008}.

With the SZ detected clusters in hand, cluster redshifts are found through optical follow-up. The mass selection function is determined with a combination of SZ, optical (cluster-member spectroscopy and weak lensing), and X-ray measurements for a subset of the clusters, allowing mass-observable relations to be used for the larger sample.  The masses and catalogs enable studies of the number distribution, $dN/dz$ or $dN(>M)/dM$, which is exponentially sensitive to the dark matter and dark energy densities and serves as a potent test for the amplitude of matter fluctuations, $\sigma_8$, and the equation of state of dark energy, $w$ \cite{Majumdar_Mohr:2004,Vanderlinde/etal:2010}. The combination of wide and deep SZ surveys ($\S$\ref{sec:obs}) helps to break cosmological parameter degeneracies thus improving constraints \cite{Khedekar/etal:2010}.

\subsection {Additional Correlations with Low Redshift Data}
As the Universe evolves, galaxies and clusters of galaxies emerge from the primordial plasma
and leave their imprint on the CMB. The CMB can be related to the foreground objects through scattering (SZ, kSZ, and reionization effects) or through gravitational lensing as described above. Additionally, by correlating CMB observations with galaxy velocity fields inferred from spectroscopic data,
one may search for the ``missing baryons" in the
outskirts of clusters \cite{ho/dedeo/spergel:2009}. In another example, ACTPol cross-correlation with massive galaxies has the potential to measure the energy feedback from supermassive black
holes, which heats the surrounding intergalactic medium and creates a
small-scale SZ distortion \cite{chatterjee/etal:2008}. It may even be possible to detect the
Kamionkowski-Loeb effect \cite{Kamionkowski1997} to see whether the quadrupole seen by observers at $z \simeq 0.3$-0.7 is similar in character to the one we observe today.
The various phenomena are distinguishable through their specific correlations and spatial distributions. The cross correlations may be done with radio, infrared, optical, and X-ray surveys, and through gravitational lensing as discussed above.

\section{OBSERVING STRATEGY}
\label{sec:obs}

\begin{figure}
\begin{center}
\includegraphics[width=5in]{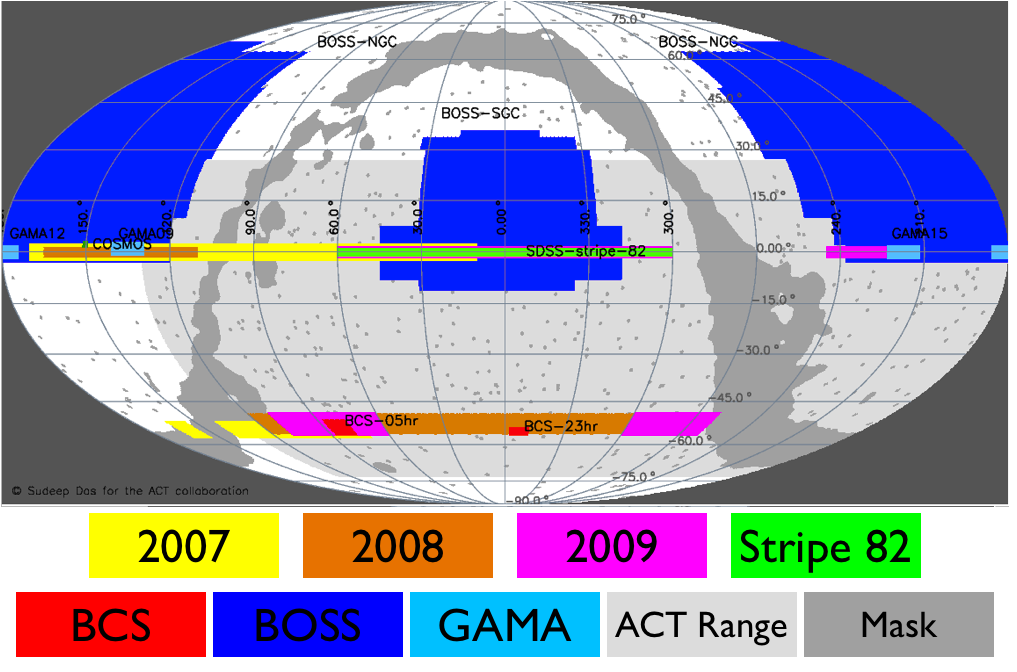}
\caption{Equal-area projection of the ACT sky coverage in celestial coordinates. The darker gray Galaxy and spots
correspond to the WMAP galactic and point source mask.  The light gray shows the $\sim$ 25,000 deg$^2$ accessible to ACT. The colored areas indicate observations by ACT during 2007-9 (yellow, orange and pink), BCS (red),
COSMOS, SDSS stripe 82 (green), BOSS (dark blue), and GAMA (light blue) surveys. The 2009 ACT coverage overlaps with the SDSS equatorial stripe. On the equator, ACT has covered 800 deg$^2$, 220 deg$^2$, and 440 deg$^2$ in 2007-9 respectively. In the southern strip, ACT has covered 280 deg$^2$, 800 deg$^2$, and 450 deg$^2$ in 2007-9, respectively. The total
BOSS coverage is 7500 deg$^2$ in the NGC and 3100 deg$^2$ in the SGC, of which ACT
can overlap 2300 deg$^2$ and 2700 deg$^2$, respectively. The region between
$130^\circ<RA<180^\circ$ corresponds to the area
that is difficult to observe with both rising and setting scans, since we typically go offline from Dec 15-Apr 1 due to poor observing conditions. This region is accessible, as can be seen in our 2007-8 equatorial coverage; it is just not optimal.}
\label{fig:coverage}
\end{center}
\end{figure}

One of the great benefits of ACT is its location. Some of the ACT science requires that we
use data from external surveys (e.g., to measure cluster redshifts, X-ray masses, and lensing masses for the SZ surveys). In other cases, the cross-correlations open up new avenues of investigation and provide alternative methods for achieving the science goals (e.g., measuring the sum of the neutrino masses).  The location of ACT enables observations of the same regions of sky covered by many other surveys. Figure~\ref{fig:coverage} shows the accessible coverage along with the coverage of some overlapping surveys.
With its elevation range of $40^\circ$ to $55^\circ$ for science observations and latitude of $-23^\circ$, ACT can observe over 25,000 deg$^2$ in $-73^\circ<\delta<27^\circ$.

Our observing strategy is based on a balance of sensitivity requirements, the need to calibrate to WMAP and Planck,  and the desire to optimize the lensing signal and the overlap with other surveys while also maximizing  the cluster science. The current temperature observations focus on two regions. The ``equatorial strip''  overlaps the SDSS  Stripe 82 and part of the BOSS  region, and somewhat less deeply the GAMA fields\footnote{The GAMA project, http://www.eso.org/\~jliske/gama/ }. It is one of the best studied regions of sky. The other, the ``southern strip,"  covers $-55^\circ<\delta<-49^\circ$ and overlaps the Blanco Cosmology Survey (BCS \cite{menanteau09/etal:2010}) fields (Figure~\ref{fig:coverage}).  The planned ACTPol observing strategy builds on this basis of multi-frequency coverage but implements significant contrast between ``deep" and ``wide" coverage.

\subsection{Sky Coverage and Overlap}

The first three ACTPol observing seasons will be split between a wide survey covering 4000 deg$^2$ to a target sensitivity of 20 $\mu$K/arcmin$^2$ in temperature and 28 $\mu$K/arcmin$^2$ in polarization at 150~GHz  (ACTPol Wide) and a deep survey that covers five $2^\circ \times 15^\circ$ regions with a target sensitivity of 3 $\mu$K/arcmin$^2$ in temperature and 4 $\mu$K/arcmin$^2$ in polarization at 150~GHz (ACTPol Deep).

The wide fields are designed for maximal overlap with rich multi-wavelength imaging data and spectroscopic data,  enabling unique multi-wavelength science.  With the wide fields we measure the growth rate of structure, the neutrino mass, curvature, and early-time dark energy, as discussed in \S\ref{sec:science}.

The ACTPol Wide field will overlap with the BOSS survey. BOSS will carry out a redshift survey of 1.5 million LRGs over 10,000 deg$^2$ and 160,000 QSOs over
an 8,000 deg$^2$ footprint\footnote{Additionally, SDSS has 600,000 quasars with photometric redshifts
\cite{richards/etal:2009}, and 15,000 MaxBCG clusters \cite{Koester2007}, and  NVSS \cite{Condon1998} has identified 400,000 1.4 GHz sources in the ACTPol Wide field.}.
ACTPol is projected to detect $\sim1000$ clusters with
$M_{\odot}>5 \times 10^{14}$ in this field \cite{sehgal/etal:2010,Tinker/etal:2008}. These clusters will have a median redshift of $\sim 0.4$, and so should significantly overlap with the SDSS cluster catalogs.

In addition to overlapping with BOSS, we are collaborating on a project to conduct a wide  (2000~deg$^2$-8000~deg$^2$) optical lensing survey of the equatorial fields with HyperSuprime Camera (HSC), a 1.5$^\circ$ FOV camera on the Subaru telescope.  The data will be optimized for weak lensing and will include measurements in the near-infrared Y band to detect $z > 1$ galaxies.

The deep regions are designed especially to optimize the signal-to-noise on the polarization and temperature and simultaneously allow for detailed studies of cluster masses. They will be spaced roughly 5 hours apart so that there are always two deep fields available for observation. We have tentatively selected the fields to overlap regions with deep optical, infrared, and X-ray observations  (XMM-LSS), and the Herschel Multi-tiered Extragalactic Survey (HERMES).

HSC also plans a deep 300~deg$^2$ survey in five optical bands ($g,r,i,z,y$) and has agreed to work to maximize the overlap between the ACT and HSC survey regions. When HSC has mapped a significant fraction of the ACTPol Deep survey, we will be able to identify photometric counterparts for the observed clusters and measure their masses.

\subsection {Foreground Emission}

Although interesting in its own right, foreground emission from diffuse sources such as
dust and synchrotron emission in our Galaxy and from radio/IR point sources outside the Galaxy complicates some cosmological analyses.

For temperature anisotropy, in the regions of the sky selected by
ACT, 150~GHz emission is contributed by both flat-spectrum radio
sources (generally the brighter 150~GHz sources) and by IR sources
(generally fainter but more numerous \cite{Vieira/etal:2009}).  Our top-level approach
to reduce point-source contamination in ACTPol images is to clean the 150~GHz data
by use of the 220~GHz data and surveys of radio sources conducted at
lower frequencies (such as the 20 GHz survey of Murphy {\it et al.} \cite{Murphy/etal:2010}).  Ultimately the degree of cleaning will limit what can be learned at 150~GHz from total intensity (i.e., unpolarized) maps. The point sources should not have a large effect in the cross-correlation studies.

While foregrounds dominate the CMB temperature fluctuations at $\ell > 2500$, CMB polarization fluctuations should dominate foregrounds sources out to $\ell \sim 5000$ (Figure~\ref{fig:pol}). The dominant foreground source above $\sim100$~GHz is expected to be dusty galaxies, which are only weakly polarized   (1~\%-2~\%) \cite{seiffert/etal:2006,tucci/etal:2005}, while the CMB is highly polarized (17~\%) at  $l > 1000$.  The cleanest frequency for polarization is near 150~GHz \cite{dunkley/etal:2008}. Synchrotron emission is a significant contaminant even at 90~GHz and is much more diffuse than the dust, which concentrates in the galactic plane \cite{hinshaw/etal:2009}.
Radio point sources are polarized at the 1~\%-5~\% level \cite{Battye/etal:2010} and have already been cleaned from the ACT maps to a depth sufficient for a polarization analysis.  Smith {\it et al.} \cite{smith2008} report that the impact of point sources on lensing is negligible.

\subsection {Calibration}

As the CMB field matures, a connected set of calibrations is developing.  WMAP has calibrated the CMB to 0.2~\% accuracy. Calibration of the planets useful to ACT (Mars, Saturn, Uranus) at the few-percent level is well underway\cite{hill/etal:2009}.  For polarization angle calibration, the primary method for ACTPol will also be to map astrophysical sources that have been characterized by other instruments.  For example, the polarization angle of Tau-A was measured by WMAP to better than $1^\circ$ \cite{Weiland/etal:2010}.  Sub-orbital instruments and Planck will continue this process, improving the temperature calibrations and measuring polarization angles with higher precision.

\section{THE ATACAMA COSMOLOGY TELESCOPE}
\label{sec:telescope}

\subsection{Telescope Overview}

ACT is an off-axis Gregorian telescope with a six-meter projected aperture primary mirror and is located on the Atacama Plateau, Chile at 5190 meters  elevation. The size of the primary mirror was driven by the requirement to obtain arcminute resolution at the ACT frequencies. The off-axis Gregorian configuration of the primary and two-meter secondary mirrors provides an unobstructed image of the sky. A detailed description of how the telescope and receiver optical parameters were optimized is presented in Fowler {\it et al.}\cite{fowler/etal:2007}.  The current receiver on ACT is the Millimeter Bolometer Array Camera (MBAC) \cite{swetz/etal:2008,thornton/etal:2008}, which includes three bolometer arrays operating at 148 GHz, 215 GHz, and 277 GHz \cite{niemack/etal:2008,zhao/etal:2008}.  The telescope control systems and mechanical performance are described in Switzer {\it et al.}\cite{switzer/etal:2008} and Hincks {\it et al.}\cite{hincks/etal:2008}

\subsection{Scan Strategy}

We scan the sky to separate the effects of the atmosphere and instrumental drift
from the celestial signal. For ACT, we scan the entire 40 metric ton upper
structure $\pm$3.5$^\circ$ in azimuth while holding the elevation
fixed (typically at 50.5$^\circ$). We move the telescope beam on the
sky on timescales faster than the $1/f$ knee of the low frequency
noise but more slowly than the time constants of the detectors.  Scans are done at two positions
east and west of an arc between the south celestial pole (SCP) and the
zenith.  As the sky rotates, an annulus around the SCP is mapped out.

This observing strategy has several benefits for observations from the ground.  First, changing the amount of air mass affects the gain of the system from atmospheric absorption.
Keeping the elevation fixed thus reduces scan-synchronous atmospheric signals.
Second, by performing both east and west  scans, the
mapped annulus will be observed in two different cross-linked orientations.
Cross-linking has been shown to be highly beneficial for
the removal of scanning-induced systematic effects such as striping
from the maps\cite{bennett/etal:2003, tegmark:1997b}.  Finally, by moving
the entire upper structure of the telescope, including the primary,
secondary, and receiver, the detectors are constantly looking through
the same optical chain.  Scanning with the complete optical system on
the sky (as opposed to using an optical chopping mirror) avoids many
scan-synchronous signals that could potentially arise from changing
the optical path such as beam shape, mirror emission, and ground
pick-up.

\section{THE ACTPOL RECEIVER}
\label{sec:receiver}

The ACTPol receiver is a cryogenic instrument designed to measure both the intensity and polarization of radiation between one millimeter and three millimeters in wavelength.  Here we describe the coupling optics, detectors, and cryogenics for the receiver.

\subsection{ACTPol Optics Design}

\begin{figure} %[htdp]
\begin{center}
 \includegraphics[width=0.49\columnwidth]{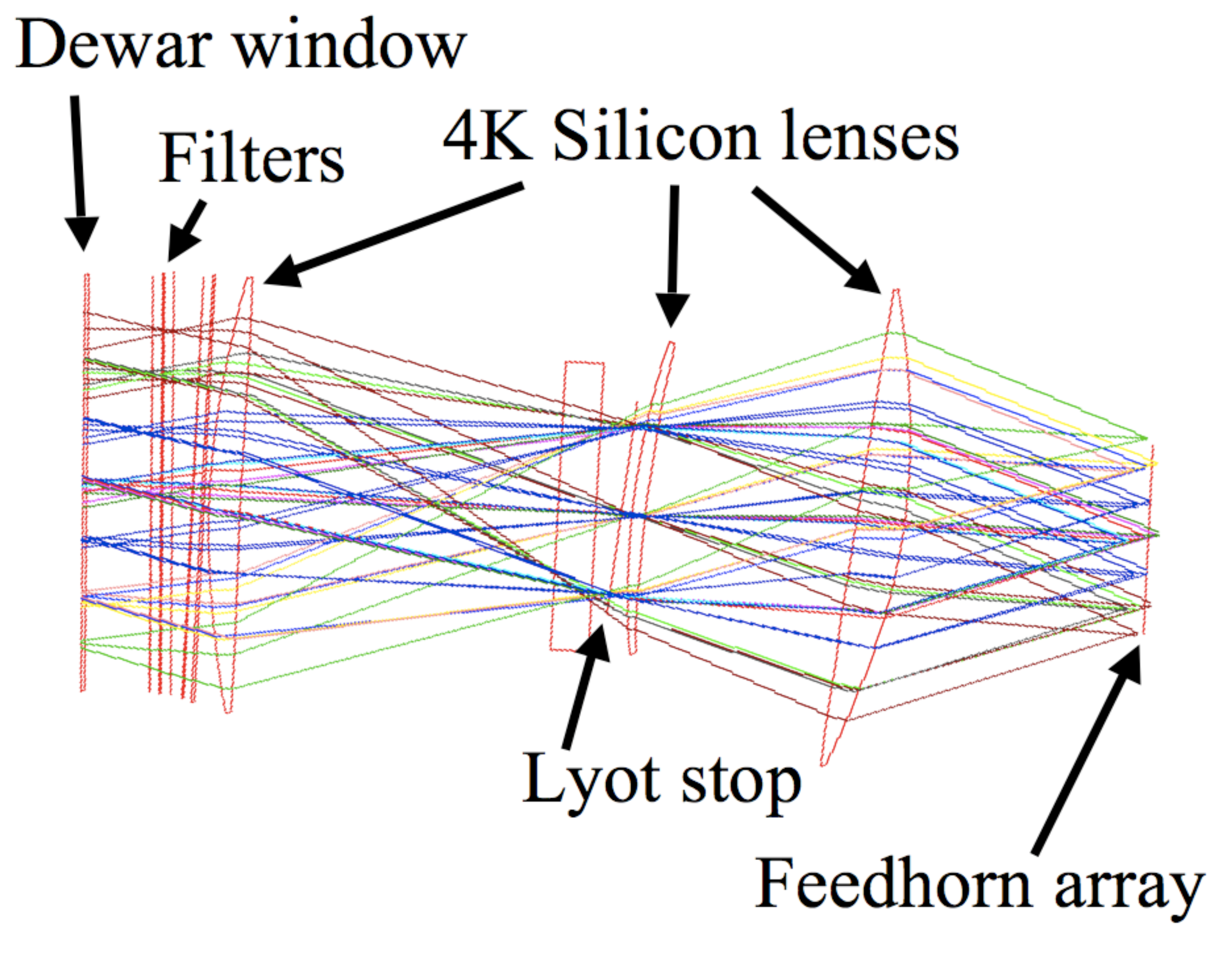}
 \includegraphics[width=0.49\columnwidth]{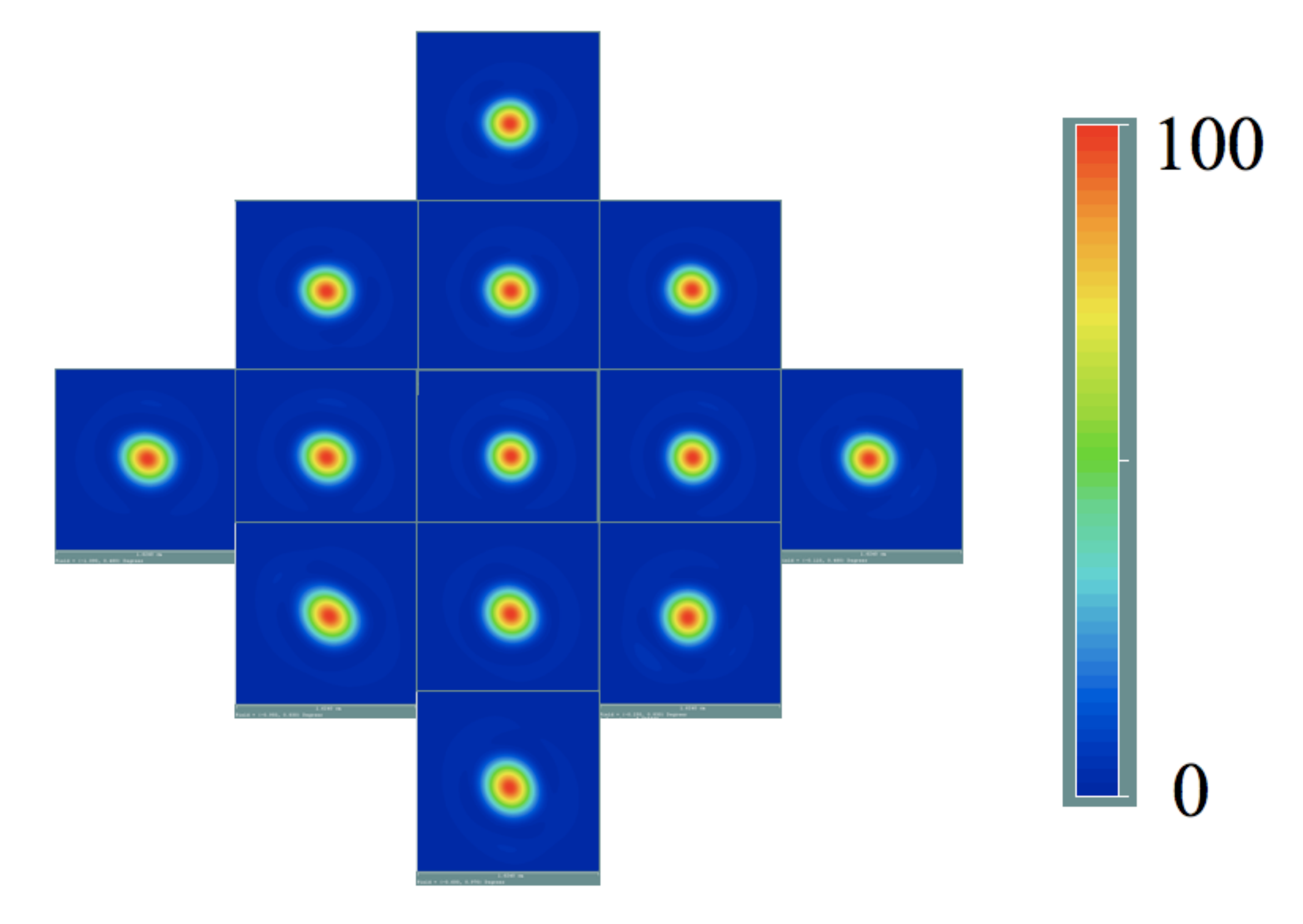}
\caption{{\it Left}: Ray-traced optical design of one of the ACTPol 150 GHz cameras. The Gregorian focus is visible near the dewar window at the left. For scale, the dewar window as drawn is 32~cm diameter.  Three silicon lenses reimage the focus onto a feedhorn array. The Lyot stop is at a primary mirror image and is used to define the primary illumination. The telecentricity is
apparent since the chief rays for different field points
are parallel to each other and perpendicular to the feedhorn array
focal plane. This design achieves Strehl ratios greater than 0.95 across the nearly one degree diameter focal plane. {\it Right}: Point spread functions (PSF) calculated for detector positions that cover the entire 150 GHz field of view (e.g., the edge PSFs shown are at the edge of the circular array).  The PSF full-width-half-maxima are $\sim1.4$ arcmin.  This design exhibits minimal beam ellipticity or asymmetry. }
\label{fig:optics_and_psfs}
\end{center}
\end{figure}

The ACTPol optics are optimized for use with new polarization-sensitive detector arrays being developed by the Truce collaboration\footnote{The Truce detector development collaboration is composed of members at NIST,  University of Chicago,  University of Colorado,  University of Michigan, NASA GSFC, and Princeton University: http://casa.colorado.edu/\~henninjw/TRUCE/TRUCE.html}.  These arrays use single-moded corrugated feedhorns to couple radiation
to the detectors ($\S$\ref{sec:det_arrays}). The refractive optics are designed to maximize the optical throughput onto three detector arrays by use of three independent optics tubes.  Two of the arrays will observe at 150 GHz and the third at 220 GHz (with the possibility of a 90 GHz replacement array).  Each optics tube uses three silicon lenses to reimage the Gregorian focus and to generate a Lyot stop, or image of the primary mirror (Figure~\ref{fig:optics_and_psfs}).  The optical throughput is limited by the maximum available size of the filters ($\sim30$ cm diameter \cite{Ade/etal:2006}), which results in a $\sim1^{\circ}$ diameter field-of-view (FOV) that is well matched to the detector arrays.  The resulting 150 GHz feedhorn aperture is $\sim 1.5 F \lambda$, where $F \approx 1.4$ is the focal ratio and $\lambda$ is the wavelength.  This spacing results in $\sim 4$ dB edge taper at the Lyot stop, or a feedhorn spillover efficiency of $\sim0.7$ through the stop.

By positioning the dewar window and the first lens near the ACT focus (Figure
\ref{fig:optics_and_psfs}), we minimize the physical size of the
optical elements; however, the increased FOV means that significantly larger optical elements are required than were used in MBAC.  Because silicon has high thermal conductivity, a high index of refraction, and low loss at our wavelengths, it is the baseline for the ACTpol optics.

The optical requirements for illuminating a flat feedhorn array are
also more stringent than for the free-space coupled bolometers
used in MBAC.  In particular, it is critical to have a
telecentric focal plane\footnote{In a telecentric focal plane the chief ray originating from the center of the illuminated region on the primary mirror is perpendicular to the focal plane surface for all positions in the focal plane.} to maintain symmetric beams and to preserve the polarization purity across the feedhorn array.

The increased throughput and telecentricity requirements of ACTPol
drive the optical design to have more shaped surfaces than MBAC in order to achieve an optimized, diffraction-limited focal plane. By making both sides of the three ACTPol
lenses curved and allowing the lenses to have small tilts and offsets,
we are able to achieve high optical quality (Strehl ratios $>$ 0.95)
across a large telecentric focal plane, as shown in
Figure~\ref{fig:optics_and_psfs}.

Two anti-reflection coatings are being considered for the ACTPol
lenses.  The first is replication of the CIRLEX coatings used in MBAC\cite{lau/etal:2006a}.
We calculate that the CIRLEX in MBAC causes $\sim15$~\% efficiency
reduction per optics tube due to loss and reflections, so we are developing simulated dielectric anti-reflection coatings as an alternative.  The simulated dielectrics are fabricated by removing much of the silicon material to a controlled depth from the surface of the lens at sub-wavelength scales, which reduces the effective index of refraction.  Three approaches are being explored to achieve this: laser-machining, deep-reactive-ion-etching, and cutting with a silicon dicing saw.  An optimized two-layer simulated dielectric coating is predicted to reduce the net loss to $<3$~\% per optics tube\cite{britton/etal:2010b}.

\begin{figure}
   \begin{center}
   \includegraphics[height=1.8in]{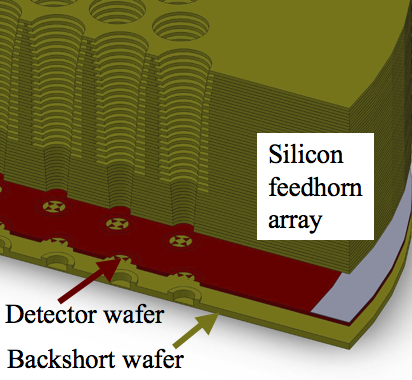}
   \includegraphics[height=1.8in]{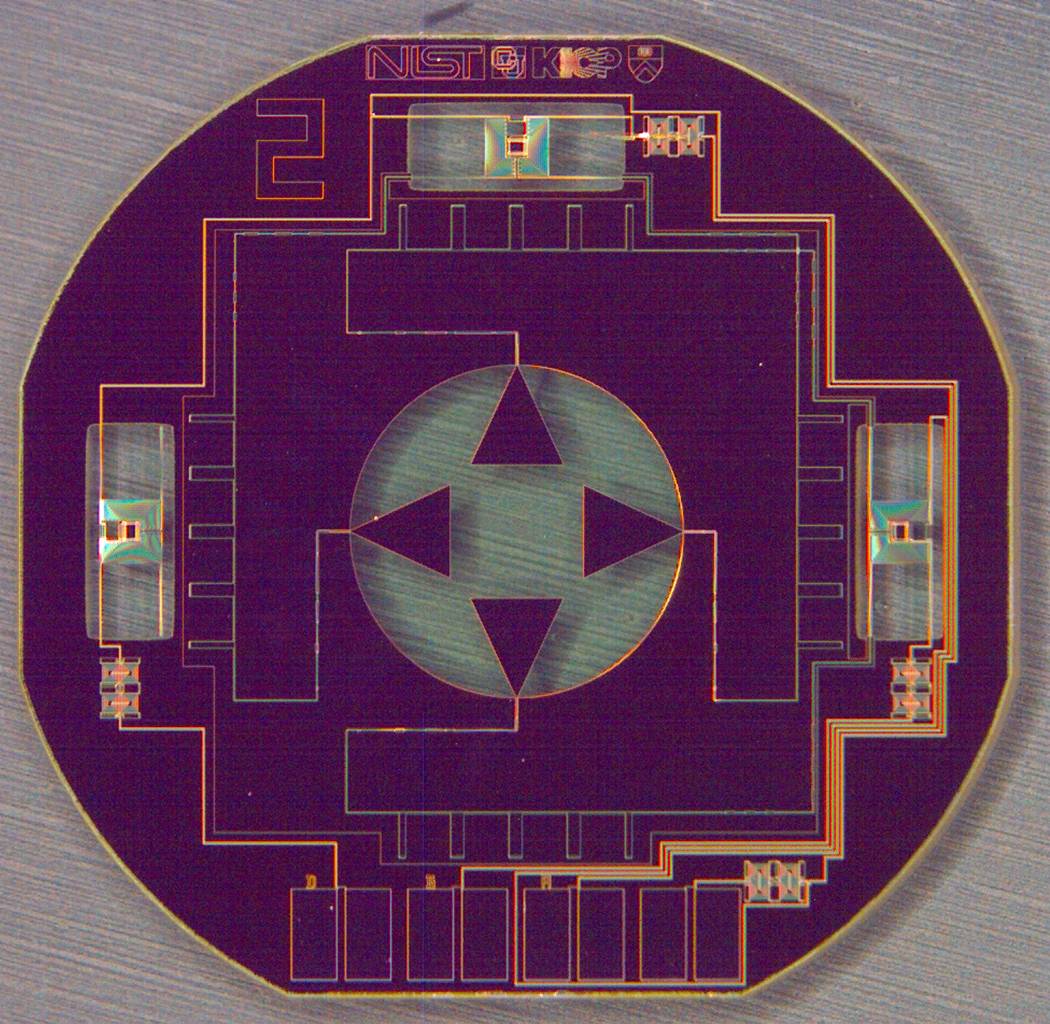}
   \includegraphics[height=1.8in]{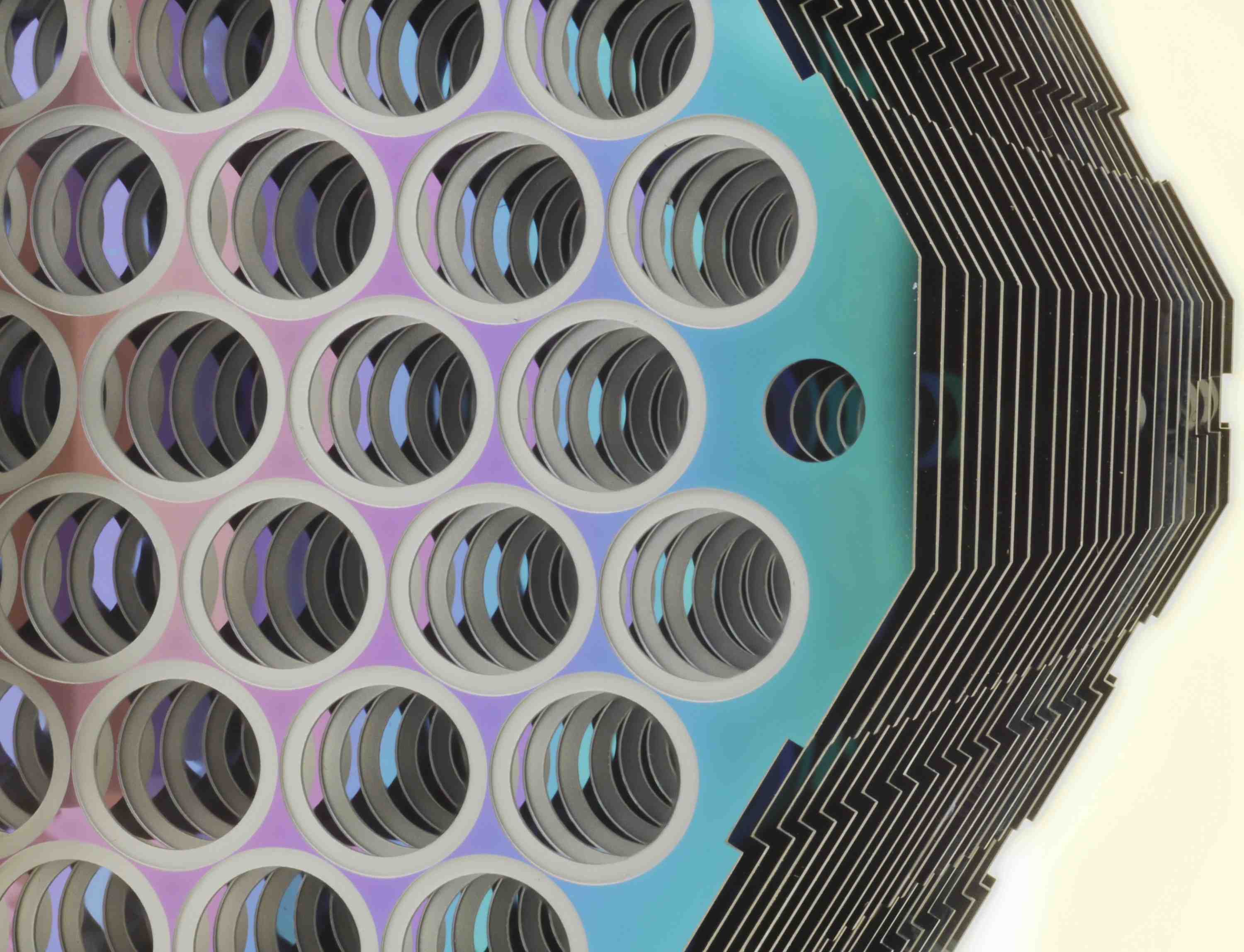}
   \end{center}
\caption{\small
{\it Left}: Cutaway view of a silicon feedhorn coupled polarimeter array design.  The spacing between the feedhorns is 5.3~mm.  The detectors and backshort wafers have been offset from the feedhorn array for clarity. {\it Middle}:  Photograph of a single prototype polarimeter that is 5~mm diameter.  The polarimeter design and performance have been described in detail\cite{yoon2009,appel2009,austermann2009b,bleem2009,mcmahon/etal:2009}. Individual polarimeters like this will be used by the Atacama B-mode Search \cite{essinger/etal:2009}, and a monolithic polarimeter array with silicon feedhorns will be used by SPTPol \cite{mcmahon/etal:2009b}. {\it Right}: Etched silicon wafers for a prototype corrugated silicon feedhorn array \cite{britton/etal:2010a} that have been offset from each other for visual clarity. The wafers will be gold plated and bonded together into a monolithic feedhorn array before detectors are installed behind them. }\label{fig:pixel}
\end{figure}

\subsection{Detector Arrays}
\label{sec:det_arrays}

The ACTPol detectors couple to incoming radiation via
monolithic corrugated feedhorn arrays coated with gold (Figure~\ref{fig:pixel}). Currently in
development at NIST\cite{britton2009,britton/etal:2010a}, the feedhorn arrays will be assembled from individual 15~cm diameter
silicon wafers with micro-machined circular apertures that correspond to
individual corrugations. The silicon wafer platelets will then be stacked, bonded and
gold-plated to form a close-packed feedhorn array.
This approach preserves the advantages of corrugated feedhorns (low sidelobes, low cross-polarization, wide-band
performance, etc.) while reducing the difficulty of
building such a large array by use of traditional techniques
(direct machining or electroforming individual metal feeds).
While similar arrays can be made from metal platelets, considerations
of weight, thermal mass, alignment, and differential thermal contraction with the
detector arrays present difficulties that can be circumvented with the
silicon platelet array concept.

A prototype 150 GHz polarimeter is depicted in Figure~\ref{fig:pixel}. The number of polarimeters that can fit into each array is limited to $\lesssim 650$ by the minimum practical spacing of each polarimeter ($\sim 5$ mm) and the maximum usable diameter of the micromachined silicon feedhorn array ($\sim 14$ cm). The polarimeters use superconducting transition-edge sensor (TES) bolometers to measure changes in radiation power.  When the TES is appropriately voltage-biased, electrothermal feedback (ETF) maintains it at the transition temperature, $T_c$, under a wide range of observing conditions \cite{irwin1995,lee1996}.   Its  steep resistance-vs-temperature curve transduces  temperature fluctuations into current fluctuations, which are read out by use of sensitive SQUID amplifiers in the time-division multiplexing architecture developed at NIST \cite{chervenak1999, dekorte2003} and implemented with room-temperature electronics provided by UBC \cite{battistelli2008}.  The ETF speeds the response of the TES to temperature fluctuations compared to its natural thermal time constant.  Each of the two TESes per polarimeter (one for each linear orthogonal polarization) is  a molybdenum-copper bilayer with normal resistance $R_n \approx 6~\mbox{m}\Omega$, which absorbs power from a  pair of microstrip transmission lines  emanating from a planar orthomode transducer \cite{mcmahon/etal:2009}  (OMT) and terminated in lossy gold meanders on the TES's thermally-isolated island.  The island connects to the bulk, at temperature $T_b$,  through four long legs composed of silicon nitride membrane topped with niobium wiring and sandwiched in silicon dioxide that have a total thermal conductivity $G$.
Microstrip filters \cite{bleem2009} define the bandpasses.   Gold resistors on the island are used as heaters that serve as responsivity calibration transfer standards and allow {\it in situ} measurements of the detector time constants.    Each detector requires a shunt resistor ($R_s \approx  180~\mu\Omega$)  for voltage biasing and an inductor  ($L \approx 1~\mu$H) to lowpass-filter the current fluctuations.

Since April 2008, the Truce collaboration has designed, produced and tested three generations of fully functional polarimeters operating at 300~mK \cite{yoon2009}.  As reported in Bleem {\it et al.} \cite{bleem2009} we have measured the detector bandpass, the detector efficiency and the leakage between the orthogonal modes ($<$~3~\%).  In Austermann {\it et al.}\cite{austermann2009b} we report on the wafer-wide uniformity of $T_c$ ($\pm 1$~\%), $R_n$ ($\pm 10$~\%), $G$ ($\pm 10$~\%), and  the detector time constants, $\tau$ ($\pm 10$~\%).  We have modeled the TES by fitting to its complex impedance data and then predicted its noise performance to within 10-20~\%\cite{appel2009}.  We are now building the first prototype polarimeter array with $\sim 100$ feedhorns and detectors \cite{britton/etal:2010a,yoon2010}.

\subsection{Sensitivity Forecast}

The baseline plan for the ACTPol receiver is to cool the detector arrays to a bath temperature of $T_b \approx 100$~mK rather than 300~mK (as for MBAC).   The lower temperature reduces the detector thermal noise below the photon noise. The latter is proportional to the loading;  a detector sensitive to only one linear polarization mode receives half the loading of an unpolarized detector, which increases the payoff for reducing the thermal noise.

To achieve our science goals we target the map sensitivities indicated in \S\ref{sec:obs} for the ACT Wide II and Deep surveys.  We introduce several conservative assumptions in calculating the detector sensitivities.  Although we predict the yield on the installed detectors will be 90~\% or higher, for the calculations here we assume 70~\%. We forecast realizing the same effective number of hours on the CMB per month that we achieved with MBAC in 2008. We make conservative (low) estimates of the optical efficiencies of the receiver based on measurements of components in hand, rather than using our predictions for improvements (in the detector efficiency, and possibly in the lenses);  similarly we use the measured lower limit on the detector bandpass, which is 30~\% smaller than the design.  These considerations lead to a goal for the noise equivalent power from thermal fluctuations in the detector of $\mbox{NEP}_G \lesssim 2.0\times 10^{-17}~\mbox{W}/\sqrt{\mbox{Hz}}$. Tests at NIST and Princeton\cite{appel2009, austermann2009b}  on prototype detectors indicate that their noise performance is straightforward to predict from their thermal conductance and is sufficiently low (but is not yet optimized for the expected ACTPol optical load or the target bath temperature).

\begin{figure}
  \begin{center}
      \includegraphics[width=5.0in]{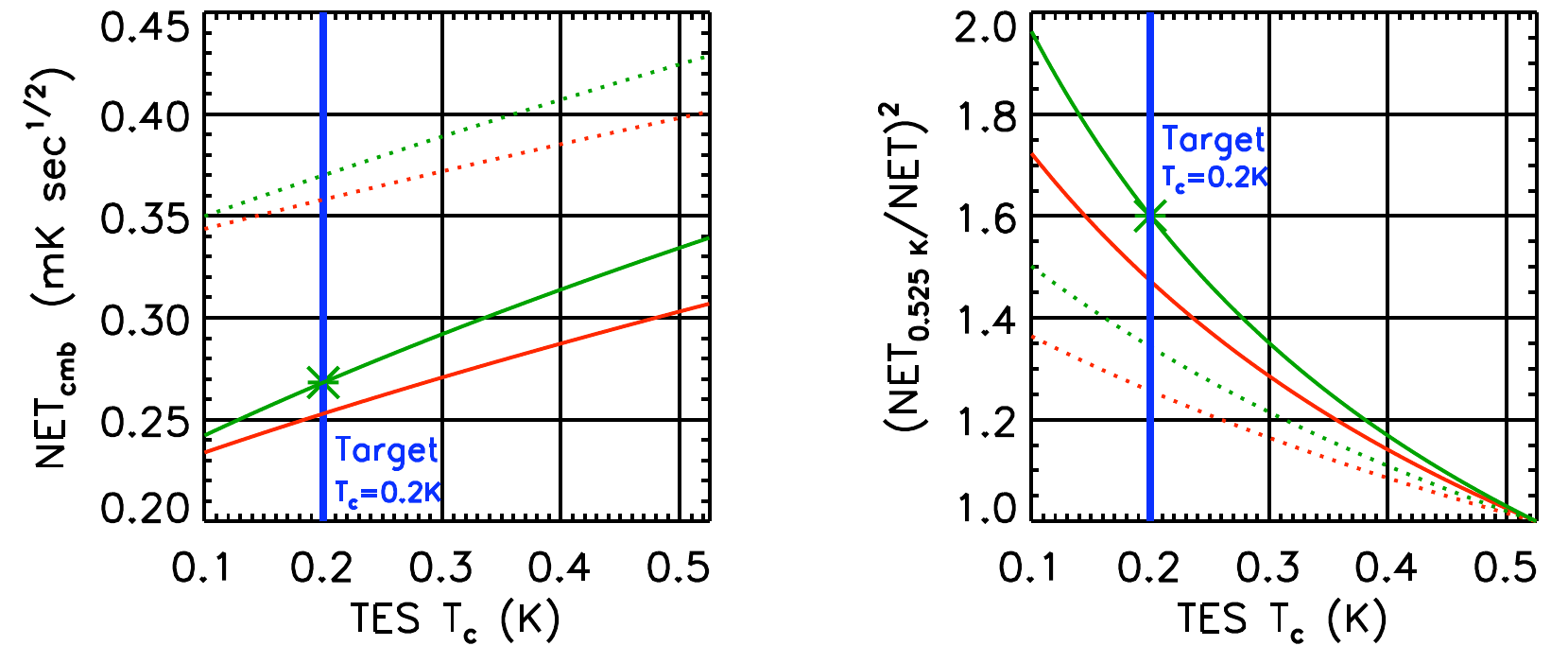}
  \end{center}
\caption{\small  Improvement with 100 mK  vs 300 mK cryogenics for a 150 GHz detector.  {\it Left:}   the sensitivity of each TES bolometer to CMB temperature fluctuations under load as a function of the critical temperature $T_c$ of the TES.  {\it Right:}  the improvement in mapping speed as  $T_c$ decreases;  the abscissa gives the ratio of the weight after a fixed period observing with $T_c=0.525$~K (as for $\sim$ 300~mK cryogenics) to that from lower $T_c$.  Here we have assumed $T_c = 5/3 T_{bath}$ and optimized $G$ for the $P_{sat}$ target.   In each plot, the green (red) curves are for $P_{sat} = 10$ $(7)$~pW, while the solid (dotted) curves are for 0.5 (2.0)~mm of precipitable water vapor (PWV).   The blue line indicates the targeted $T_c = 0.2$~K, and the green asterisks indicate the sensitivity assumed for the error estimates in $\S$\ref{sec:science}.  We will choose the  $P_{sat}$ target after our initial prototyping runs of detectors with $T_c = 0.2$~K;  for now we conservatively estimate $P_{sat}\approx 5P_\gamma$, where $P_\gamma$ is the total photon load.  The median PWV for the 2008 ACT observing season was $0.5$~mm.  }
\label{fig:det_noise}
\end{figure}

We determine the target parameters for $T_c$ and $G$ consistent with the noise and detector time-constant goals.  Our model for optimizing these parameters includes radiative loading and noise from the CMB, the atmosphere, the warm ACT mirrors, the dewar window,  and cryogenic elements, as well as the detector noise, saturation power, $P_{sat}$, and bath temperature, $T_b$ \cite{zmuidzinas2003}.  Figure~\ref{fig:det_noise} illustrates how the NEP varies with $T_c$ for two sets of atmospheric loading conditions and two $P_{sat}$ values. The right panel shows the improvement in mapping speed from operating at $T_b \approx 100$~mK as compared to $\sim300$~mK.  These calculations and the estimated number of working detectors lead to the target ACTPol instrument noise equivalent temperatures of 6 $\mu$K $\sqrt{\mbox{s}}$ at 150 GHz and 20 $\mu$K $\sqrt{\mbox{s}}$ at 220 GHz.

\subsection{The ACTPol Cryostat}
The cryo-mechanical aspect of the ACTPol receiver will be similar to, but larger than, the existing MBAC receiver.
The primary cooling for the receiver will be provided by two pulse tube refrigerators.
Because the bandpasses for the ACTPol bolometers are defined on the detector wafers rather than by free-space filters, we need only cool the optical elements to 4~K.  Additional cooling stages are then required to cool only the detector arrays and to provide  thermal isolation.  The baseline plan for the cryostat includes three such stages. The first stage will be cooled to $\sim$ 0.8 K by a $^4$He sorption refrigerator similar to the ones in  MBAC. The second is a 300~mK $^3$He sorption refrigerator, which will back single-stage adiabatic demagnetization refrigerators (ADRs) that cool the detector arrays to 100~mK. We predict $> 30$ hour hold times for the ACTPol He fridges and $> 50$ hour hold times for the ADRs, which are more than two and three times longer than needed for observations.

\begin{figure}
   \begin{center}
   \includegraphics[width=3.3in]{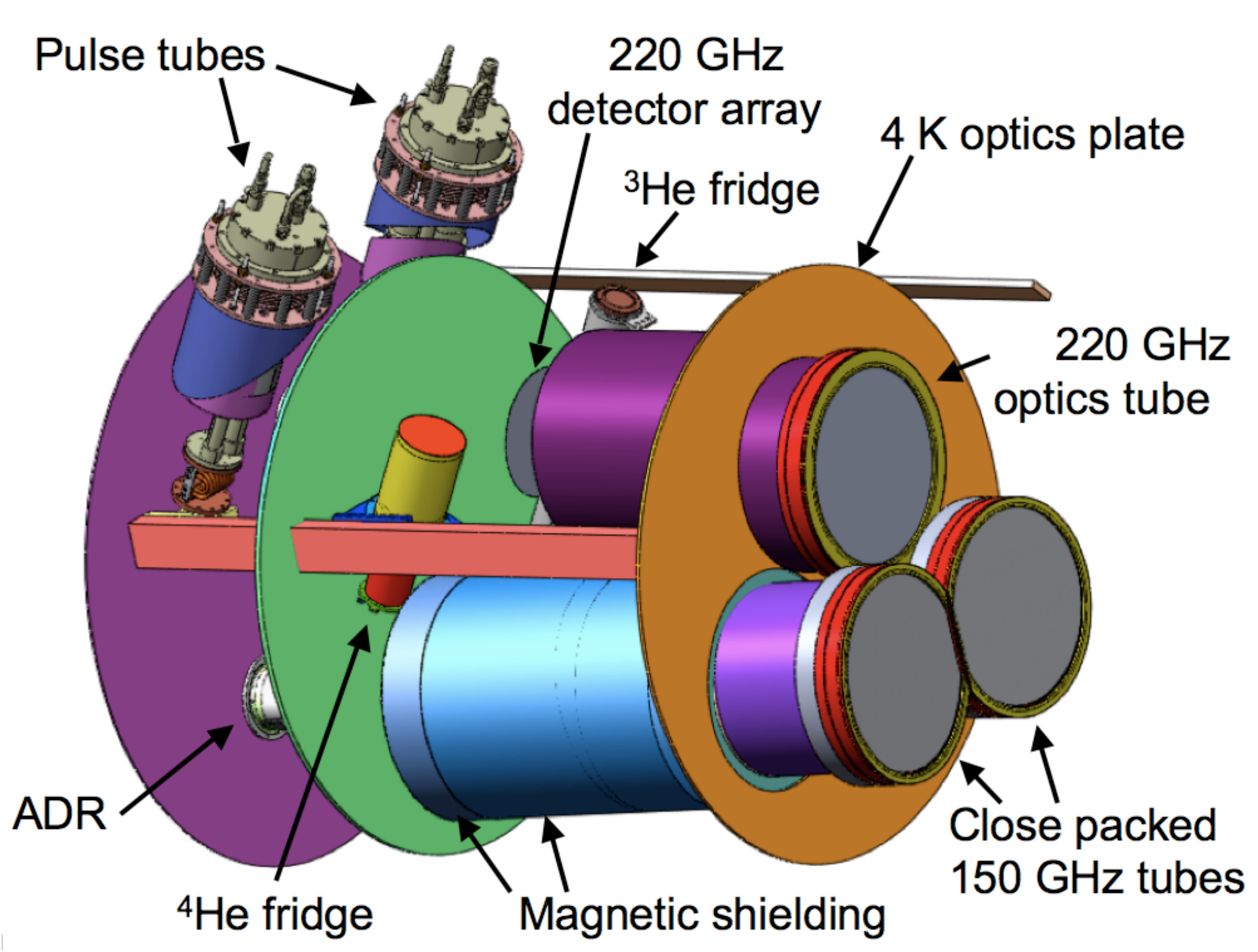}
   \includegraphics[width=3.3in]{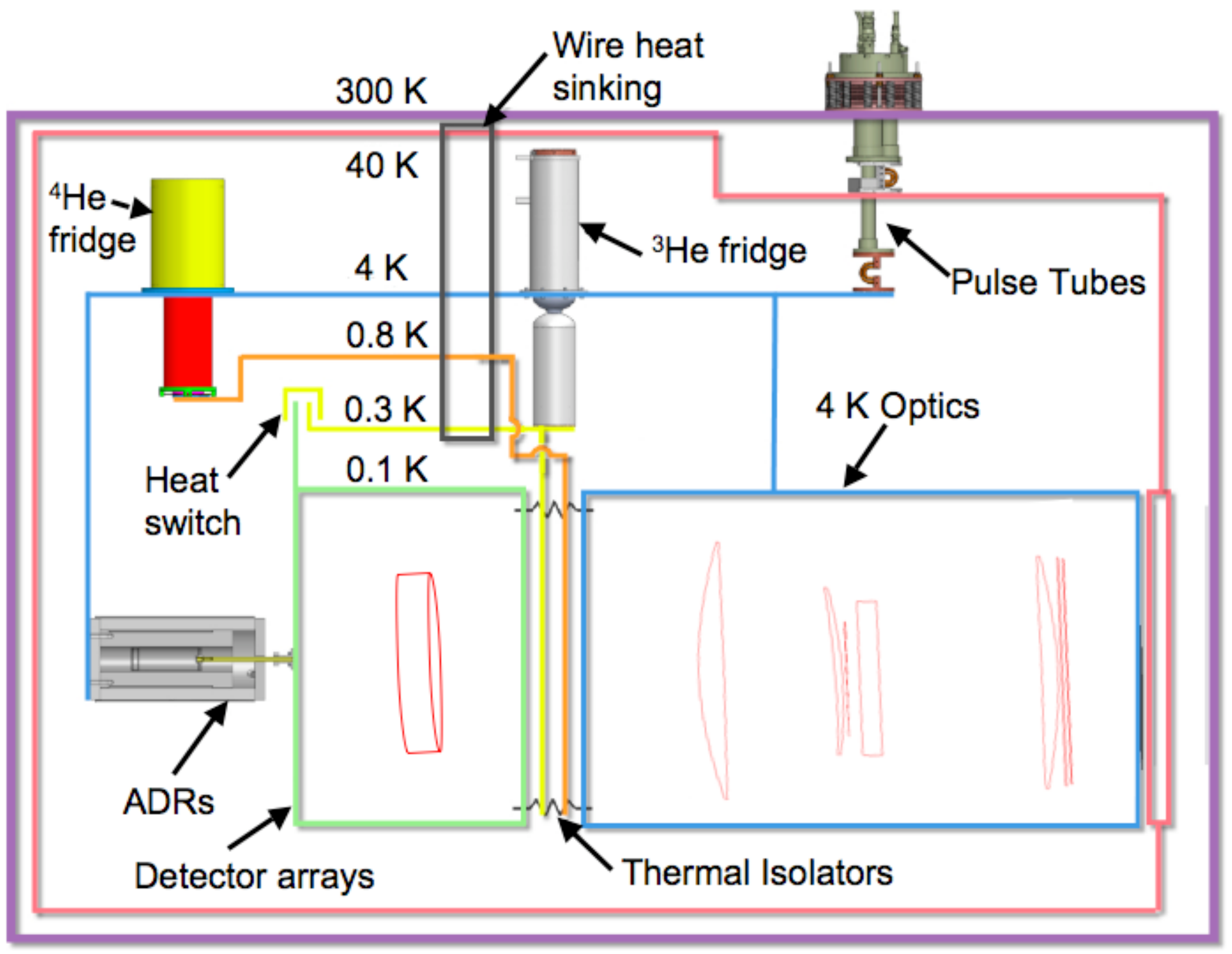}
   \end{center}
   \caption
   { \label{fig:Receiver_cryo}
{\it Left}: Rendering of a preliminary ACTPol receiver design.  The entrance of each optics tube is $\sim30$~cm diameter. {\it Right}:  Schematic of cryogenic links in the receiver.}
\end{figure}

\subsection{Improvements in the ACTPol Receiver Compared to the MBAC Receiver}

The ACTPol receiver is a considerable upgrade over MBAC.   The most obvious improvement is that these detectors have polarization sensitivity that enables the new science described earlier. The colder operating temperature allows for lower NEPs over what is possible with 300~mK cryogenics. The ACTPol field-of-view and optical throughput are both substantially larger than that of MBAC.  We also are baselining two detector arrays, instead of one, at the critical science band of 150 GHz.  In addition to having more detectors, this means there will be improved overlap of 150 GHz and 220 GHz observations on the sky.

Another substantial improvement is that we project the ACTPol detectors to have better noise performance than those in MBAC, which had excess in-band noise.  The prototype ACTPol detectors  exhibit in-band NEPs consistent with thermal fluctuation noise for $0.5 < F_{link} < 1$ \cite{austermann2009b}, whereas the MBAC detectors suffer from a median value of roughly 60~\% excess in-band noise compared to the model with $F_{link} = 1$ \cite{niemack:2008}.

These combined improvements in detector NEP, optical throughput, number of detectors, and bath temperature result in the target ACTPol sensitivity being more than four times better than that of the current MBAC receiver for 150 GHz temperature measurements, in addition to its polarization sensitivity.

\section{CONCLUSION}

The development of the ACTPol receiver is underway, while CMB temperature observations with the existing MBAC receiver continue.  ACTPol will observe both wide and deep fields, overlapping with a variety of observations accessible from the Atacama and enabling a wide range of science goals, including: measuring the gravitational lensing of the CMB, constraining the sum of the neutrino masses and key parameters of inflation, and cross-correlating SZ and lensing measurements with other surveys. Target first light for ACTPol with three detector arrays is 2013.

\acknowledgments

We thank members of the ACT and Truce collaborations for critical contributions to this project. ACT is supported primarily by the U.S. National Science Foundation through awards AST-0408698 for the ACT project, and PHY-0355328, AST-0707731 and PIRE-0507768. CMB detector development at NIST is supported by the NIST Innovations in Measurement Science program.  Funding was also provided by Princeton University and the University of Pennsylvania.  Niemack was supported by a National Research Council Postdoctoral Fellowship.

%%%%%%%%%%%%%%%%%%%%%%%%%%%%%%%%%%%%%%%%%%%%%%%%%%%%%%%%%%%%%

\bibliography{report}   %>>>> bibliography data in report.bib
\bibliographystyle{spiebib}   %>>>> makes bibtex use spiebib.bst

\end{document}